\documentclass[journal,transmag]{IEEEtran}

\makeatletter
\DeclareRobustCommand*\textsubscript[1]{%
  \@textsubscript{\selectfont#1}}
\def\@textsubscript#1{%
  {\m@th\ensuremath{_{\mbox{\fontsize\sf@size\z@#1}}}}}
\makeatother

\usepackage[pdftex]{graphicx}
\usepackage{amsmath, amssymb}
\usepackage{lipsum}
\usepackage[noadjust]{cite}
\usepackage{physics}
\usepackage{xcolor}

\begin{document}

\title{Reversal Time of Jump-Noise Dynamics for Large Nucleation}

\author{\IEEEauthorblockN{Arun Parthasarathy\IEEEauthorrefmark{1} and
Shaloo Rakheja\IEEEauthorrefmark{1}
\IEEEauthorblockA{\IEEEauthorrefmark{1}Department of Electrical and Computer Engineering,
New York University, Brooklyn, NY 11201, USA}
\thanks{
Corresponding author: A. Parthasarathy (email: arun.parth@nyu.edu).}}}

\IEEEtitleabstractindextext{%
\begin{abstract}
The jump-noise is a phenomenological stochastic process used to model the thermal fluctuation of magnetization in nanomagnets. In this work, the large nucleation regime of jump-noise dynamics is studied, and its reversal time is characterized from Monte Carlo simulations and analysis. Results show that the reversal time of jump-noise dynamics for large nucleation is asymptotically equal to the time constant associated with a single jump-noise scattering event from the energy minimum in the energy landscape of the magnetization. The reversal time for large nucleation depends linearly on the height of the energy barrier for large barriers.
The significance of the large nucleation regime of jump-noise dynamics to phenomenologically explain the magnetoelectric switching of antiferromagnetic order parameter is also prospected.
\end{abstract}

}

\maketitle
\IEEEdisplaynontitleabstractindextext
\IEEEpeerreviewmaketitle

\section{Introduction}

\IEEEPARstart{T}{he} reversal time of magnetization is the time constant associated with {the} longitudinal relaxation of magnetization in nanomagnets under the influence of thermal effects, also known as the superparamagnetic or N\'eel relaxation time~\cite{coffey2012thermal}. The reversal time provides information on the retention time for reading or the switching speed for writing in the context of a magnetic memory or a logic device~\cite{kryder2008heat, yang2013memristive}. Estimating the reversal time correctly is crucial for steady miniaturization of such devices against the onset of thermal instability in smaller volumes.

The jump-noise~\cite{mayergoyz2011magnetization} is a phenomenological stochastic process used to model {the} thermal fluctuation of magnetization in nanomagnets.
Unlike {the} classical N\'eel-Brown thermal activation theory~\cite{brown1963thermal, brown1979thermal} based on coherent rotation of magnetization via Landau-Lifshitz-Gilbert dynamics~\cite{Landau:437299, gilbert2004phenomenological}, the magnetization reversal via jump-noise dynamics~\cite{parthasarathy2018reversal} represents macroscopic tunneling of magnetization~\cite{tejada1993quantum}, a low temperature (10 mK--10 K) escape rate phenomenon, without evoking quantum mechanics~\cite{liu2012jump}.

Besides being used to capture thermal effects in magnetization dynamics, the jump-noise could also model the magnetoelectric (ME) switching of antiferromagnetic (AFM) order parameter~\cite{martin1966antiferromagnetic, belashchenko2016magnetoelectric, kosub2017purely}, the emerging field of antiferromagnetic spintronics~\cite{baltz2018antiferromagnetic}.  In AFMs such as Cr\textsubscript{2}O\textsubscript{3}, the ME energy density required for domain switching is nearly four orders of magnitude
smaller than the uniaxial anisotropy energy barrier~\cite{belashchenko2016magnetoelectric}. In such materials, classical thermal activation over the energy barrier, which depends exponentially on the barrier height, cannot explain the broad range of fast switching speeds reported in the literature from microsecond~\cite{toyoki2015magnetoelectric} to few tens of nanosecond~\cite{kosub2017purely} to few tens of picosecond~\cite{manipatruni2015spin, nikonov2014benchmarking}. The jump-noise in the large nucleation regime is expected to phenomenologically address this anomalous switching by virtue of it being a first-order phase transition model nucleated by fluctuations~\cite{langer1968theory}.
Our prior work shows that the dynamics of jump-noise averages to the classical theory when the nucleation is small~\cite{parthasarathy2018reversal}. Hence, the case of small nucleation is only briefly presented for the sake of completeness. 

In this paper, the jump-noise and the large nucleation regime are first defined (Sec.~\ref{sec:jump-noise}).
Then, the reversal time extracted from Monte Carlo simulations of jump-noise dynamics in the large nucleation regime (Sec.~\ref{sec:tauMC}) is compared with that obtained from analysis (Sec.~\ref{sec:RD}).

\section{The Jump-Noise}
\label{sec:jump-noise}
For a uniformly magnetized nanomagnet with magnetization $\vb{M}$ of magnitude $M_\text{s}$, the state variable is defined by the dimensionless quantity $\vb{m} = \vb{M}/M_\text{s}$.
The jump-noise is characterized by the transition probability rate function $S$ between any two states $(\vb{m}_1, \vb{m}_2)$ on the phase space $\norm{\vb{m}}=1$, which is given by the formula~\cite{mayergoyz2011magnetization}
\begin{align}
S(\vb{m}_1,&\vb{m}_2) = B\exp\bigg[-\frac{1}{2\sigma^2}\norm{\vb{m}_1-\vb{m}_2}^2 + \nonumber \\ &\frac{e_{\text{b}0}}{2}\big\{g(\vb{m}_1)-g(\vb{m}_2)\big\}\bigg]; \quad e_{\text{b}0} = \frac{\mu_0M_\text{s}^2V}{kT},
\label{eq:S}
\end{align}
where $B$ and $\sigma$ are the nucleation parameters; $g$ is the magnetic free energy density; and $e_{\text{b}0}$ is the energy barrier parameter, wherein $\mu_0$ is the vacuum permeability, $V$ is the volume of the nanomagnet, and $kT$ is the thermal energy.
From Eq.~(\ref{eq:S}), the scattering rate $\lambda$ from a state $\vb{m}$ follows 
\begin{equation}
\lambda(\mathbf{m}) = \oint_{\norm{\vb{m}'}=1} S(\vb{m}, \vb{m}')\dd[2]{m'}.
\label{eq:lambda}
\end{equation}
The probability density function $f$ of a jump to occur from $\vb{m}_i$ to $\vb{m}_{i+1}$ at time $t_i$ is written as
\begin{equation}
f(\vb{m}_i, \vb{m}_{i+1} | \vb{m}_i) = \frac{S(\vb{m}_i,\vb{m}_{i+1})}{\lambda(\vb{m}_i)}. 
\label{eq:fi}
\end{equation}
The statistic of the jump instants $t_i$ is given as
\begin{equation}
\text{Pr}(t_{i+1}-t_{i}>\tau) = \exp\left(-\int_{t_i}^{t_i+\tau} \lambda(\mathbf{m}(t)) dt\right). 
\label{eq:ti}
\end{equation}

Equations \eqref{eq:S}--\eqref{eq:ti} describe the jump-noise statistics. The nucleation regime is decided by the parameter $\sigma$. A small nucleation is such that the transition probability rate from a state is appreciable only over an infinitesimal distance from the state, implying that $\sigma\ll 1$. A large nucleation means that states which are farthest apart on the phase space have a non-negligible transition probability rate. On the unit sphere phase space, the transition probability rate between diametrically opposite states, if energetically favorable, is at least equal to $\exp(-2/\sigma^2)$. Therefore, $\sigma\sim\mathcal{O}(1)$ for large nucleation.

\section{Extraction of Reversal Time}
\label{sec:tauMC}
The reversal time of magnetization is extracted by performing Monte Carlo simulations on the time evolution of jump-noise induced nucleation in nanomagnets. The nanomagnets possess only uniaxial anisotropy, and there is no external applied field. The simulations are implemented in \mbox{MATLAB} with the help of Parallel Computing Toolbox on a server with 20-core CPU @ 2.3~GHz and 512~GB memory. The numerical methods are presented in Ref.~\cite{parthasarathy2018reversal, lee2012monte}. 

\begin{figure}[!t]
\centering
\includegraphics[width=\linewidth]{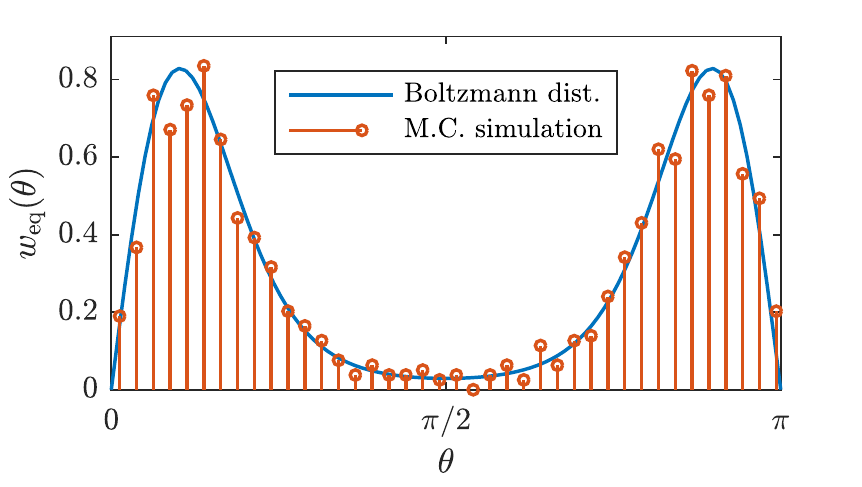}
\caption{Equilibrium distribution of $\theta$ for uniaxial anisotropy; $e_{\text{b}0} = 10$. The histogram of $\theta$ obtained from Monte Carlo simulation of jump-noise dynamics eventually approaches Boltzmann distribution.}
\label{fig:eqdist}
\end{figure}

The state variable $\vb{m}$ is represented by spherical coordinates $(\theta,\phi)$ such that $m_x = \sin\theta\cos\phi$, $m_y = \sin\theta\sin\phi$, $m_z=\cos\theta$. We consider 1000 samples aligned along the same lowest energy state $m_z=-1$, without loss of generality, at time $t=0$, and let the ensemble evolve with time until the ensemble equilibrates to Boltzmann distribution as shown in Fig.~\ref{fig:eqdist}. The energy density for uniaxial anisotropy is $g(\theta) = (1/2)\sin^2\theta$, and  the corresponding Boltzmann distribution is
$ w_\text{eq}(\theta) = (1/Z)\exp[-e_{\text{b}0} g(\theta)]\sin\theta$,
where $Z$ is the normalization constant.

The reversal time  $\tau$ characterizes the longitudinal relaxation of absolute ensemble mean of the state variable as
\begin{equation}
\abs{\overline{m_z}}(t)  \approx e^{-t/\tau}; \quad t\gg\tau.
\label{eq:Neel}
\end{equation}
So, $\tau$ can be estimated from the asymptotic value of $\tau(t) = -t/\ln[\abs{\overline{m_z}}]$ from simulations. The reversal time extracted this way will have some error because a finite sample size of 1000 could only allow precision upto three significant digits.

\section{Results and Discussion}
\label{sec:RD}
For very large nucleation, that is $\sigma\gg 1$, the first term in the transition probability rate~\eqref{eq:S} vanishes, so that
\begin{equation}
S(\vb{m}_1,\vb{m}_2) \simeq B\exp\left[\frac{e_{\text{b}0}}{2}\big\{g(\vb{m}_1)-g(\vb{m}_2)\big\}\right].
\label{eq:Slargenuc}
\end{equation}
The transition probability between two equivalent energy wells in this case is symmetric about the location of the energy barrier, so that the average distance of jumps occurs at the energy maximum. As a result, the critical process of relaxation between two equivalent energy wells against the energy barrier happens in a single random process. 

The time constant of a process is associated with the slowest mode of relaxation. From Eq.~\eqref{eq:Slargenuc}, it is evident that transitions originating from the energy minimum have the lowest escape rate, which is expected from statistcal mechanics. Therefore, the longitudinal relaxation of magnetization for very large nucleation is characterized by jump-noise scattering process from the energy minimum. The reversal time is simply reciprocal of the scattering rate \eqref{eq:lambda} from the energy minimum or formally
\begin{equation}
\tau \simeq \frac{1}{\lambda_0},\ \text{where}\ \lambda_0 = \lambda[\vb{m} = \mathrm{arg}\min{g(\vb{m})}].
\label{eq:tau}
\end{equation}

\begin{figure}[!t]
\centering
\includegraphics[width=\linewidth]{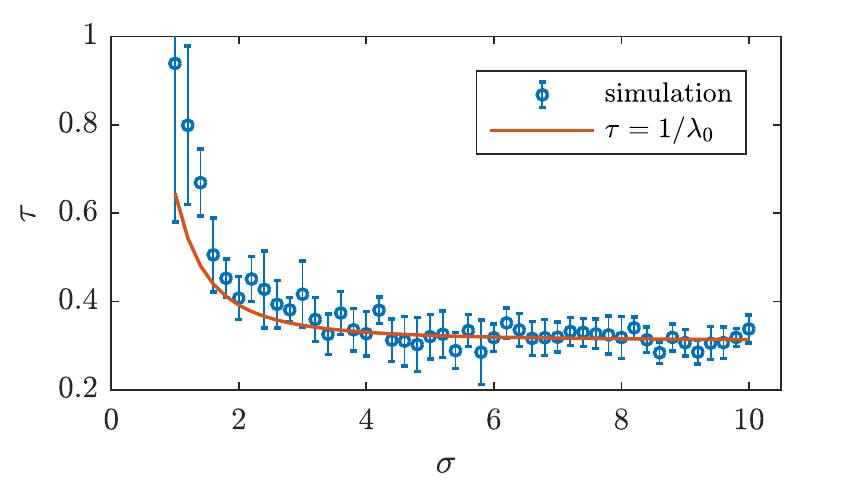}
\caption{Reversal time of magnetization for uniaxial anisotropy for large nucleation; $e_{\text{b}0} = 10$, $B=1$. The reversal time obtained from Eq.~\eqref{eq:tau} and from Monte Carlo simulations asymptotically converge for large $\sigma$.}
\label{fig:taurev_ln}
\end{figure}

The reversal time obtained from Eq.~\eqref{eq:tau} asymptotically converges to the reversal time extracted from Monte Carlo simulations for large $\sigma$ as shown in Fig.~\ref{fig:taurev_ln}. For larger values of $\sigma$, the statistics of the jump-noise process reduces to Eq.~\eqref{eq:Slargenuc} which is independent of $\sigma$, so the reversal time saturates. At $\sigma\sim 1$, there is a sharp rise in the reversal time which could be used to model critical phenomena. 

\begin{figure}[!t]
\centering
\includegraphics[scale=1]{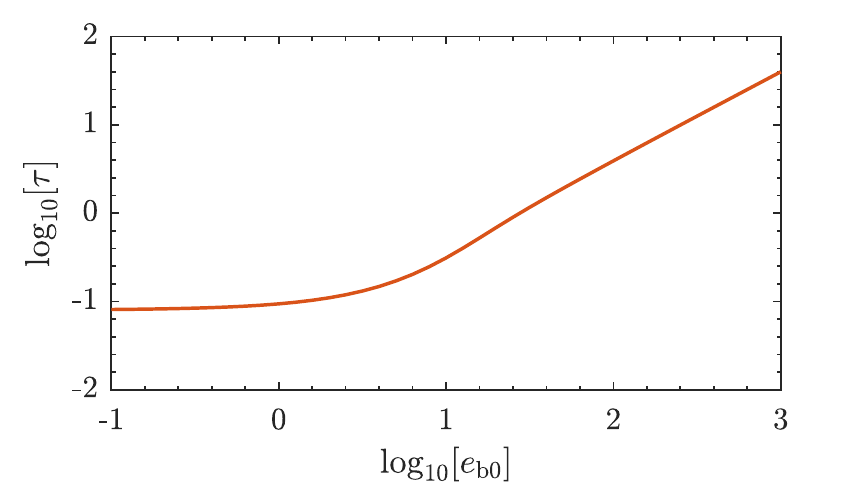}
\caption{Reversal time of magnetization for uniaxial anisotropy for large nucleation; $\sigma=100$, $B=1$. The reversal time varies linearly with the energy barrier for large $e_\text{b0}$ and saturates to $\tau=1/(4\pi B)$ for small $e_\text{b0}$.}
\label{fig:taurev_eb0}
\end{figure}

The reversal time for large nucleation varies linearly with the energy barrier for large $e_\text{b0}$ as shown in Fig~\ref{fig:taurev_eb0}. When $e_\text{b0}\gg 1$, the transition probability rate~\eqref{eq:Slargenuc} behaves like a Dirac delta function centered at the energy minimum, and the normalization of the Dirac delta yields a linear $e_\text{b0}$ term in the expression of the reversal time~\eqref{eq:tau}. For $e_\text{b0}\ll 1$,  the transition probability rate~\eqref{eq:Slargenuc} is uniformly equal to $B$ on the phase space, and the reversal time $\tau = 1/(4\pi B)$.

\begin{figure}[!t]
\centering
\includegraphics[width=\linewidth]{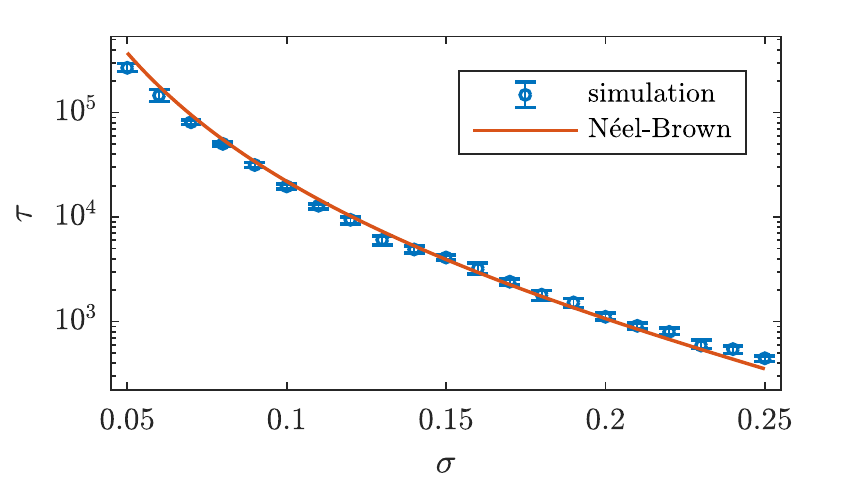}
\caption{Reversal time of magnetization for uniaxial anisotropy for small nucleation; $e_{\text{b}0} = 10$, $B=1$. The reversal time obtained from the classical N\'eel-Brown theory and from Monte Carlo simulations converge for small $\sigma$. See Ref.~\cite{parthasarathy2018reversal} for details on how N\'eel-Brown theory's solution was obtained.}
\label{fig:taurev_sn}
\end{figure}

In contrast, for the case for small nucleation, the jump-noise dynamics averages to the classical N\'eel-Brown theory~\cite{parthasarathy2018reversal}. As a result, the reversal time varies exponentially with $\sigma$ as shown in Fig.~\ref{fig:taurev_sn}, as well as exponentially with the energy barrier~\cite{coffey2012thermal}. In this regard, the jump-noise dynamics for large nucleation exhibits a unique feature.

As mentioned in the introduction, the classical N\'eel-Brown theory cannot explain the fast switching speeds of AFM domain via ME effect. We theorize that the critical phenomenon of ME switching can be explained by modeling the nucleation parameter $\sigma$ as a function of the ME energy density or the product of the electric and magnetic fields. When the field product is below the threshold, $\sigma$ and the nucleation should be small, and as a result the reversal is probabilistically suppressed because of the large energy barrier due to anisotropy. At the threshold field product, $\sigma\sim\mathcal{O}(1)$ and the nucleation is large; consequently the reversal is more favorable, despite the large barrier. 

\section{Conclusion}
The reversal time of jump-noise induced magnetization dynamics for large nucleation is asymptotically equal to the time constant associated with a single scattering event from the energy minimum. The reversal time for large nucleation depends linearly on the energy barrier for large barriers. This is in stark contrast with the classical N\'eel-Brown thermal activation theory, where the reversal occurs coherently over the energy barrier in infinitesimally many steps, and shows exponential dependence on the energy barrier. In future work, the large nucleation regime of jump-noise dynamics will be used to phenomenologically model the magnetoelectric switching of antiferromagnetic order parameter, an otherwise impossible phenomenon to explain classically.

\section*{Acknowledgment}

This work was supported in part by the Semiconductor Research Corporation (SRC) and the National Science Foundation (NSF) through ECCS 1740136. S. Rakheja also acknowledges the funding support from the MRSEC Program of the National Science Foundation under Award Number DMR-1420073.

\bibliographystyle{IEEEtran}
\bibliography{myref}

\end{document}